\documentclass[3p,times]{elsarticle}

\usepackage{amssymb}
\usepackage{graphicx}

\newcommand{\be}{\begin{equation}}
\newcommand{\ee}{\end{equation}}
\newcommand{\ba}{\begin{eqnarray}}
\newcommand{\ea}{\end{eqnarray}}
\newcommand{\ban}{\begin{eqnarray*}}
\newcommand{\ean}{\end{eqnarray*}}
\newcommand{\nn}{\nonumber}

\begin{document}

\begin{frontmatter}

\title{No Pain, No Gain: Hard Probes of the Quark-Gluon Plasma Coming of Age}

\author{Berndt M\"uller}

\address{Department of Physics, Duke University, Durham, NC 27708-0305, USA}

\begin{abstract}
This lecture presents a concise review of the current status of hard QCD and electromagnetic processes as probes of the quark-gluon plasma.
\end{abstract}


\end{frontmatter}

% \linenumbers

\section{ Introduction}
\label{sec:intro}

The manifesto of the hard QGP probes program was aptly summarized by C.~Louren{\'c}o and H.~Satz in their preface to the proceedings of the first Hard Probes conference in 2004 \cite{Lourenco:2005qp}:
``The aim of high energy heavy ion collisions is to produce and study a medium of deconfined quarks and gluons in the laboratory. {\em Deconfinement} occurs when the density of quarks and gluons becomes so high that long range confining forces cease to be effective. It {\em is thus intimately connected to short spatial scales, and to resolve and study phenomena at such scales, hard probes are essential and have to be developed into as precise tools as possible.} Hence it is necessary to study the production of heavy flavours and quarkonia, of jets, and of photons and di-leptons in strongly interacting media.'' 
The questions to be asked at the opening of Hard Probes 2012 are:  Where are we on the path toward this goal? What have we learned? What is still missing? What needs to be done? I will attempt to answer these questions in my lecture.

Hard probes in the sense of this program are: high-$p_T$ partons and jets, heavy quarks and open flavor hadrons, heavy quarkonia ($J/\psi$, $\Upsilon$, and their excited states), and electroweak probes (lepton pairs, photons, the $Z$ and $W^\pm$). The production rates for these probes are rigorously calculable in the standard model, although there are some {\em caveats}: the production of heavy quarkonia requires additional modeling, experimentally determined nuclear PDFs must be used as input, etc.  Another common feature of hard probes is that their final-state interactions can be factorized from their production; results obtained in $A+A$ collisions can be normalized to $p+p$ and/or $p(d)+A$ data. Of course, final state interactions are negligible for the electroweak probes.

The methodology applied in the use of hard probes of for QCD matter are:
\begin{itemize}
\setlength{\itemsep}{0pt}
\item
Formulate the production in $A+A$ as hard QCD process with factorizable final state interactions (FSI).
\item
Formulate FSI in terms of medium properties (e.~g.\ transport coefficients) that can be calculated for any medium model.
\item
Identify observables that are sensitive to certain aspects of the structure of the medium, e.~g.\ a weakly versus a strongly coupled QGP;  the scale separating weak from strong coupling; the possible quasiparticle structure of the QGP.
\item
Calculate medium properties relevant to FSI on the lattice, if possible.
\end{itemize}

Which properties of hot QCD matter can we hope to determine with the help of hard probes and by means of which observables? In the order of difficulty of calculating these properties by means of lattice QCD, the most prominent ones are: (1) The inverse screening length (Debye mass) of chromo-electric fields
\be
m_D = - \lim_{|x|\to\infty} |x|^{-1} \ln \langle E^a(x)E^a(0) \rangle
\ee
which affects the binding of heavy quarkonium states and causes their ``melting'' when increased beyond a certain threshold; (2) the electromagnetic current-current correlator
\be
\Pi^{\mu\nu}_{\rm em}(Q) = \int d^4x\, e^{iQx} \langle j^\mu(x)j^\nu(0) \rangle,
\label{eq:PiQ}
\ee
which determines the QGP radiance of virtual and real photons; and the transport coefficients
\ba
\hat{q} = \frac{4\pi^2 \alpha_s C_R}{N_c^2-1} \int dy^- \langle F^{a+i}(y^-) F_i^{a+}(0) \rangle ,
\nn \\
\hat{e} = \frac{4\pi^2 \alpha_s C_R}{N_c^2-1} \int dy^- \langle i \partial^- A^{a+}(y^-) A^{a+}(0) \rangle ,
\ea
which control the radiative and collisional energy loss of energetic partons. $\hat{q}$ measures the transverse momentum broadening per unit path length of a fast parton; $\hat{e}$ measures the energy transferred to the medium via collisions per unit path length.

It is worthy to note that, apart from $\Pi^{\mu\nu}$ all medium properties are expressed as correlators of color
gauge fields, i.~e.\ they reflect the gluonic structure of the QGP. At high $Q^2$ and/or high $T$, asymptotic freedom ensures that the QGP is weakly coupled and has a quasi-particulate structure. The questions we would like to answer are: At which $Q^2$ (or $T$) does the QGP become strongly coupled? Does it still contain quasiparticles? Can we use hard partons to locate the transition? Which observables tell us where the transition occurs?

As an example how hard probes could work, consider the relation between the shear viscosity $\eta$ and $\hat{q}$ proposed by Majumder {\em et al.} \cite{Majumder:2007zh} who argued that $\eta/s$ and $\hat{q}$ are related at weak coupling in gauge theories by the general relation $\eta/s \approx {\rm const.} \times T^3/\hat{q}$. On the other hand, at strong coupling, $\eta/s$ saturates at $1/4\pi$, but $\hat{q}/T^3$ increases without limit. Can this be shown to constitute and unambiguous criterion for weak vs. strong coupling? Similarly, the collisional energy loss parameter $\hat{e}$ is sensitive to the mass $m$ of scattering centers in the QGP and goes to zero in the $m\to\infty$ limit, unless the scatterers have a dense spectrum of excited states, akin to atoms. Thus $\hat{e}$ serves as a probe of medium structure at the color screening scale.

\section{Bulk evolution of the QGP}
\label{sec:bulk}

An important development relevant to the hard probes program is that the global properties of relativistic heavy ion collisions are no longer the limiting factor. It is now firmly established that the bulk evolution of the QGP can be quantitatively described by second-order viscous relativistic hydrodynamics. The equation of state of the QGP is now known from lattice-QCD calculations with sufficient accuracy, and the shear viscosity of the QGP can be inferred from measurements of the elliptic flow $v_2$, both as function of $p_T$ and as function of the centrality, with sufficient precision to render the uncertainties for the calculation of FSI of hard probes negligible. The remaining uncertainties concern the initial time $\tau_0$ of thermalization of the QGP (probably $\tau_0 < 1$ fm/c) and the precise shape of the transverse profile of the initial energy density $\varepsilon$.

A precision analysis \cite{Song:2010mg} of the centrality dependence of the elliptic flow in Au+Au collisions at the top RHIC beam energy has yielded the constraint $1 \leq 4\pi\eta/s \leq 2.5$, with the transverse energy density profile being the main source of uncertainty. Viscous hydro calculations of the event-by-event elliptic flow $v_2(p_T)$ and the triangular flow $v_3(p_T)$ for Au+Au at RHIC \cite{Schenke:2010rr} and Pb+Pb at LHC \cite{Schenke:2011tv} agree well with the data for values of $\eta/s$ in the same range. The $p_T$-spectra of identified hadrons agree very well with the RHIC and LHC data within the range of momenta ($p_T < 2-3$ GeV/c) where the hydrodynamical model can be expected to apply.

It is thus no longer a meaningful approach to draw conclusions from comparisons of hard probes calculations with data when using schematic models of the QGP bulk evolution. The days when theorists could get away with deducing a value for $\hat{q}$ or related parameters by substituting an effective path length in a static QGP for a realistic evolution model are over. This is real progress, but also a new challenge for theorists working on hard probes of the QGP.

One other fact worth mentioning is that beautiful data from RHIC and LHC of $p_T$-spectra of identified hadrons are unambiguously telling us where the ``hard'' part of the $p_T$-spectrum begins. By comparing baryon-to-meson ratios for hadrons with similar quark content (e.~g.\ the $\Lambda/K^0_s$ ratio \cite{Belikov:2011zz}) with predictions of hydrodynamic and recombination model calculations \cite{Fries:2003fr}, the onset of the truly ``hard'' part of the spectrum can be located to $p_T \geq 6$ GeV/c.

\section{Jets}
\label{sec:jets}

Another game changing development in the field of hard probes is that it is now possible to fully reconstruct jets even in the high-background environment of central Au+Au and Pb+Pb collisions. While the first attempts of full jet reconstruction at RHIC were reported at the previous Hard Probes conference by PHENIX and STAR, the much higher energy of jets available at LHC, together with the superb calorimetry of ATLAS and CMS, is opening up a new era of jet quenching measurements in heavy ion collisions. The first spectacular new observable accessible in Pb+Pb collisions at LHC is the dijet asymmetry $A_J = (P_{T1}-P_{T2})/(P_{T1}+P_{T2})$, where $P_{T1}$ ($P_{T2}$) is the total transverse momentum of the leading (subleasing) jet within a chosen jet cone. In addition, we have data for the single jet observable $R_{\rm CP}^{\rm (jet)}$ and the familiar single- and double-hadron observables $R_{AA}$ and $I_{AA}$.

A new analysis of the direct photon $R_{AA}$ data from PHENIX \cite{Afanasiev:2012dg} has shown that photons are unsuppressed in central Au+Au collisions up to the highest measured $p_T\approx 10$ GeV/c. Photon data from CMS up to $p_T=60$ GeV/c show that the same is true in central Pb+Pb collisions at LHC \cite{Chatrchyan:2012vq}. The fact that $Z$ and $W$ boson production in Pb+Pb at LHC is also unsuppressed \cite{Chatrchyan:2011ua} tells us less about the lack of FSI (the $Z$ and $W$ decay long before a QGP can be formed) than about lack of a modification of the nuclear (anti-)quark PDF in the relevant $x$-range. The single non-photonic electron data from PHENIX,\cite{Adler:2005xv} as well as the data on single $D$-mesons from ALICE \cite{Rossi:2011nx} show that mesons containing open charm are as much suppressed as light hadrons for $p_T > 5$ GeV/c. It will be interesting to see whether this conclusion remains valid when higher statistics data from LHC and data with vertex identification from RHIC become available.

Curiously, $R_{\rm CP}^{\rm (jet)}$ data from ATLAS \cite{Angerami:2012zz} show no $E_T$ dependence within errors over the range 100 -- 300 GeV; nor do they show any clear evidence for a medium modification of the jet fragmentation function in central and semi-central Pb+Pb collisions compared with peripheral events. The latter conclusion is confirmed by data from CMS \cite{Yilmaz:2011zz}. It must be noted, however, that the LHC data have a lower $p_T$-cut of 4 GeV/c, which means that  they would miss a medium induced modification if it were constrained to lower $p_T$. Also note that ATLAS and CMS define the fragmentation variable $z$ with respect to the final measured jet energy, not with respect to the (unknown) energy of the parton that initiated the jet. The result may simply mean that the jets hadronize after emerging from the medium and that any missing energy shows up at $p_T < 4$GeV/c. They are thus compatible with data from STAR that show a strong modification of the jet fragmentation functional lower $p_T$.

Theorists hoping to explain these data or to extract physics insights from them will increasingly have to rely on Monte-Carlo event generators for jet fragmentation together with jet finding algorithms. This will pose increasing demands on the sophistication of theoretical calculations and the ability of theorists to read the ``fine print'' in experimental papers. It also opens the door to a serious discussion about where the interface between theoretical calculations and experimental analysis should be.

Some theorists (e.~g.\ T.~Renk \cite{Renk:2012wi}) have argued that triggered few-particle observables like $R_{AA}$ and $I_{AA}$ are at least as discriminating, if not more so, as probes of energy loss mechanisms (because of the infrared sensitivity of jet finders) than fully reconstructed jets. Is this really true? Even if the argument applies to the present approaches, does it have to be true, or can jet reconstruction be made ``transparent''? Or will ``hard probe'' jet quenching physics degenerate into comparing theoretical MC's with experimental MC's? This clearly deserves serious discussion at this conference and in the future.

\section{The opaque QGP}
\label{sec:opaque}

A explained in the the Introduction, QCD offers two mechanisms of parton energy loss: Direct transfer of energy to the medium via collisions and radiative energy loss via emission of gluons (for a recent review of parton energy loss theory, see \cite{Majumder:2010qh}). The strength of these two mechanisms is controlled by the transport coefficients $\hat{e}$ and $\hat{q}$. For a homogeneous medium of thickness $L$, the energy loss per unit length is given by \cite{Baier:2000mf}
\be
- dE/dx = \hat{e} + (\alpha_sN_c/\pi)\hat{q}L  .
\ee
The questions, to which one would like to extract answers from the experiments, are: What is the relative importance of these mechanisms of energy loss? How are radiative and collisional energy loss affected by the structure of the medium, e.~g.\ whether it is composed of quasiparticles, and if so, what their masses are? Can the energy loss be described by AdS/CFT inspired models with a weak-strong coupling transition at a certain scale? What happens to the lost energy and momentum?  If the energy loss is predominantly ÒradiativeÓ, how quickly does the radiation thermalize. What is its longitudinal momentum ($z$) distribution? What is its angular distribution (the jet ÒshapeÓ), i.~e.\ how much is found inside a cone of angular size $R$? How do the answers depend on the parton flavor?

The ``QGP Brick'' exercise carried out by the TEC-HQM Collaboration \cite{Armesto:2011ht} has clearly revealed the deficiencies of the analytical approaches to radiative parton energy loss and provided the motivation to construct Monte-Carlo implementations of the various energy loss schemes and to develop NLO jet quenching schemes. Has the time come to carry out a similar MC energy loss challenge, or is more theoretical development needed before such an effort makes sense? 

It is a remarkable fact that many predictions for $R_{AA}(p_T)$ in Pb+Pb at LHC based on the pQCD theory of parton energy loss and extrapolating $\hat{q}$ from RHIC were not far off the mark (see \cite{Abreu:2007kv}). In particular, the calculations predicted a rather rapid rise of $R_{AA}$ with $p_T$, because the primary parton spectrum is much flatter than at RHIC. As a result, the relative suppression of hadrons for $p_T > 10$ GeV/c is stronger at RHIC than at LHC! This is not expected in theories that consider the hard parton to be strongly coupled to the medium. On important question is how to extrapolate $\hat{q}$ from RHIC to LHC. The default assumption has generally been to scale $\hat{q}$ with the entropy density, or phenomenologically, with the final charged particle multiplicity $dN_{\rm ch}/dy$ for a given centrality window. In some simplified jet quenching schemes\cite{Betz:2012qq} this leads to an overestimate of the suppression at LHC, which can be cured by letting $\alpha_s$ run with the medium scale, giving $\alpha_s^{\rm LHC} = (0.8-0.9)\alpha_s^{\rm RHIC}$. This fits well with the observation that a similar scale dependence of $\alpha_s$, in this case with the saturation scale $Q_s$, is needed to fit the centrality dependence of the produced multiplicity in the glasma model \cite{Schenke:2012hg}.

One important consideration that may guide future jet quenching studies is that parton virtuality matters. Obviously, a hard parton can only radiate gluons if it is off-shell. In a hard QCD interaction, a parton is naturally born off-shell and only gradually comes on shell as time evolves. Roughly speaking, at a time $t$ after the primary hard scattering, the average virtuality measured in the parton's rest frame, is $Q^2(t)\approx E/L(t)$ by virtue of the uncertainty relation, where $L(t) \approx t$ is the distance traveled in the medium. The additional virtuality acquired by the parton through collisions in the medium is $\Delta Q^2 = \hat{q}L$. If $Q^2(L) \gg \Delta Q^2(L)$, the parton hardly notices the medium, i.~e.\ its radiation pattern is dominated by vacuum radiation. This is where parton energy matters \cite{Muller:2010pm}. Since $Q^2$ scales with parton energy, but $\Delta Q^2$ only with final multiplicity, a 20 GeV/c parton at RHIC with $\hat{q}L \approx 4.5~{\rm GeV}^2 \gg E/L = 1.5~{\rm GeV}^2$ is a ``better'' medium probe than a 200 GeV/c parton at LHC with $\hat{q}L \approx 9~{\rm GeV}^2 < E/L = 13~{\rm GeV}^2$.

The large increase in the fraction of dijets showing a large asymmetry $A_J$ in Pb+Pb collisions at LHC, first reported by ATLAS and confirmed by CMS, can be understood as an effect of ``jet collimation'' \cite{CasalderreySolana:2010eh}. This term describes the effect that the medium removes most low-$z$ fragmentation gluons from the jet cone, either by deflection or by energy degradation and thermalization. The collimation process can be described by a momentum diffusion equation for the gluons accompanying the primary parton of the form \cite{Qin:2010mn}:
\be
\frac{\partial}{\partial t} f_g(\omega,k_\perp,t) = \hat{e} \frac{\partial f_g}{\partial\omega} 
   + \frac{1}{4} \hat{q}\, \nabla_{k_\perp}^2 f_g + \frac{dN_g}{d\omega d^2k_\perp dt} ,
\ee
where the last term denotes the rate of medium induced gluon radiation. The experimentally measured $A_J$ distribution in central Pb+Pb collisions can be reproduced by such an approach, or similar ones, with values for $\hat{e}$ and $\hat{q}$ that are reasonable extrapolations from those describing $R_{AA}$ at RHIC \cite{Qin:2010mn,Young:2011qx,He:2011pd,Renk:2012cx,ColemanSmith:2012vr}. It thus seems that the color opacity and stopping power of the QGP produced at LHC is at least semi-quantitatively described by perturbative QCD jet quenching theory and extrapolation of the medium properties from RHIC, possibly with a slight decrease in $\alpha_s$ expected from the increase in the temperature of the medium.

\section{The screened QGP}
\label{sec:screen}

The essence of a quark-gluon plasma is, of course, the screening of chromo-electric fields due to the presence of free color charges. It is worth recalling that a plasma is defined as a globally neutral state of matter containing mobile charges. Interactions among the charges of many particles spread a localized charge over a characteristic  length (the Debye length) and thus screen the charge. In strongly coupled plasmas \cite{Ichimaru:1982zz} the Debye length is small and only a few particles exist in the Debye sphere. This results in non-negligible nearest-neighbor correlations among charges, which is an indicator of the liquid-like properties of strongly coupled plasmas.

Color screening in the QGP can be probed with heavy quark bound states. The question to be asked is, which ones survive at which temperature? While most of the experimental emphasis at SPS and RHIC has been on the charmonium states, the higher rate of $b$-quark pair production at LHC shifts the emphasis to the $\Upsilon$ states, $\Upsilon$(1s), $\Upsilon$(2s), and $\Upsilon$(3s). These constitute even better probes than the charmonium states, because (a) they are not fed from decays of heavier states (e.~g., the $J/psi$ can be produced by $b$-quark decays), and (b) heavy quark effective theory and non relativistic (NR-) QCD is much more reliable for the $b\bar{b}$ system than for the $c\bar{c}$ system. One important question is whether recombination of $c\bar{c}$-quarks contributes to the charmonium yields at RHIC and LHC \cite{Rapp:2008tf}. If it does, the effect should become clearly visible for LHC energies. A related question is whether residual correlations among heavy quark pairs, which are co-produced, can enhance the recombination probability?

The original concept of heavy quarkonium melting, proposed in \cite{Matsui:1986dk}, was simple: the $Q\bar{Q}$ bound state would ``melt'', when the color screening length would become shorter than the bound state radius. As often, the real story is more complicated, as has been recognized in recent years, because the heavy quark bound state can interact with the medium both elastically and inelastically, e.~g., by colliding with a thermal gluon. In the NRQCD limit, the wave function of a heavy quarkonium state obeys an effective Schr\"dinger equation of the form \cite{Akamatsu:2011se}:
\be
i\hbar \frac{\partial \Psi_{Q\bar Q}}{\partial t} 
= \left[ \frac{p_Q^2 + p_{\bar Q}^2}{2M} + V_{Q\bar Q} -\frac{i}{2}\Gamma_{Q\bar Q} + \eta \right] \Psi_{Q\bar Q} ,
\ee
where $\eta$ is a noise term describing the stochastic influence of the medium on the center-of-mass motion of the quarkonium state. When the decay width $\Gamma_{Q\bar Q}$ exceeds the binding energy of the $Q\bar{Q}$ state, the state effectively ``melts''. In practice, this means that the melting of the quarkonium states occurs similarly as the melting of the $\rho^0$ state near $T_c$. In a recent analysis \cite{Laine:2011xr}, the $\Upsilon$(1s) state melts at $T \approx 270$ MeV, when it is still bound by approximately 50 MeV. One important insight to take away is that the collisional energy loss of heavy quarks and the suppression of heavy quarkonia are strongly intertwined and cannot be separated. A credible theory needs to describe both processes correctly.

Using these techniques combined with a variational treatment of the viscous hydrodynamic expansion of the quark-gluon plasma, Strickland \cite{Strickland:2011mw,Strickland:2011aa} has calculated the time-integrated suppression factor $R_{AA} = \int d\tau\,\Gamma_{Q\bar Q}(\tau)$ for several $b\bar{b}$ bound states in Pb+Pb collisions at LHC energy. While some data, e.~g.\ the STAR data on inclusive $\Upsilon$ suppression as function of centrality, are described quite nicely, there are significant discrepancies with some of the CMS data on $\Upsilon$(1s) suppression. It also turns out that the remaining uncertainties in the choice of $V_{Q\bar Q}$ are large.

While much remains to be done theoretically and experimentally (we need much higher statistics data!), there has been remarkable progress since HP2012. Clearly, the suppression of the individual bund states should not be treated independently, but as part of a coupled channel calculation of all bound states, including possible coupling to the continuum describing ionization and recombination. On the experimental side, the data on $J/\psi$ suppression from SPS, RHIC, and LHC remain bewildering and defy a simple, systematic interpretation. Given the fact that several different mechanisms contribute to $R_{AA}^{J/\psi}$, this is not unexpected, but it means that only a comprehensive theoretical treatment of $J/\psi$--medium interactions, including cold nuclear matter effects, color screening, thermal dissociation, and recombination, can hope to describe the data. The good news is that the framework for such a treatment is coming within reach.

\section{The shining QGP}
\label{sec:shining}

Only the electromagnetic probes, photons and lepton pairs, carry direct information out of the interior of the QGP. Because final-state interactions for these probes are negligible, their theory is in good shape. For example, the emission rate of lepton pairs with four-momentum $Q=(\omega,{\bf q}$ is given by
\be
\frac{dN(\ell^+\ell^-)}{d^4x\,d^4Q} = \frac{e^4}{6(2\pi)^4 Q^2} (1+2m^2/Q^2)(1-4m^2/Q^2) \Pi^\mu_\mu(Q)
\ee
with $\Pi^{\mu\nu}(Q)$ given by (\ref{eq:PiQ}). As a dynamical real-time quantity, $\Pi(Q)$ cannot be calculated directly on the lattice, but its spectral density $\rho(\omega,{\bf q}=0)$ can be deduced by analytic continuation of the imaginary-time current-current correlator. Recent results of quenched lattice calculations \cite{Ding:2010ga} suggest that the spectral density at thermal frequencies is strongly enhanced above the result for free quarks, but not quite as strongly as predicted by HTL perturbation theory. This is precisely what one would expect from a strongly coupled QGP.

The direct photon data from PHENIX \cite{Adare:2009qk} obtained as extrapolation of the low-mass di-lepton spectrum to zero invariant mass shows clear evidence of thermal emission in the range $p_T < 2.5$ GeV/c. The data are in good agreement with the predictions based on a hydrodynamical expansion with early thermalization $\tau_{\rm th} \leq 1$ fm/c, thus providing direct evidence of the formation of a quark-gluon plasma with an initial temperature $T_{\rm in} \geq 2T_c \approx 300$ MeV.

\section{Conclusion}
\label{sec:summary}

In order to attain their promise, hard probes of hot QCD matter must overcome a number of challenges:

The pQCD theory of jet quenching must finally become quantitative. Theorists must extend the validation effort to include MC codes and NLO treatments of gluon radiation. The large kinematic span from RHIC to LHC will be crucial for model discrimination; where LHC and its superb high-acceptance calorimetric detectors provide superior jet reconstruction capabilities, which RHIC provides a better medium-to-vacuum virtuality match. The question, whether reconstructed jets really provide additional information compared with single- and two-hadron observables, warrants serious discussion. Similarly, it will be important to explore the sensitivity of jet transport coefficients to the medium structure; after all, we don't just want to learn something about the dynamics of jets in a dense QCD medium, but gain insight into the structure of the quark-gluon plasma itself.

A quantitative theory of the heavy quarkonia interactions with the medium is almost at hand. However, we need a better understanding what to use as the $Q\bar{Q}$ potential. Also, high statistics measurements of quarkonium suppression in $p(d)+A$ collisions will be required to isolate the hot medium contribution from initial-state effects such as gluon shadowing and cold nuclear matter suppression. For the electroweak probes the most urgent need are high statistics, low background data.

Summarizing, it is worth revisiting the Hard Probes manifesto: ÒTo resolve and study a medium of deconfined quarks and gluons at short spatial scales, hard probes are essential and have to be developed into as precise tools as possible.Ó  It is fair to state that, as of 2012, hard probes have yet to fulfill their promise:  In the case of jets and quarkonia, the investigation is mainly theory limited. Given good data, we do not yet know how to reliably extract $\hat{q}$ and $\hat{e}$.  We do not yet know which jet observables are most sensitive to the physics we want to learn.  A quantitative theory of quarkonium suppression is just emerging. In the case of photons and di-leptons, better data are needed.

But progress is being made rapidly, as HP2012 will show in abundance, and the goal appears ultimately reachable. 

{\em Acknowledgments:} This work was supported in part by grants DE-FG02-05ER41367 and DE-SC0005396 from the Office of Science of the U.~S.~Department of Energy.

%\bibliographystyle{elsarticle-num}
%\bibliography{/Users/BoMue/Documents/TEX/BMrefs,/Users/BoMue/Dropbox/ARN_Review/LHC}


\end{document}